\documentstyle[12pt,epsf,cite]{article}
\input{epsf}
\newcommand{\ds}{\displaystyle}

\begin{document}

\title{
On the Possibility of Experimental Verification\\
of Some Localization Theory Predictions}
\author{A.G.~Groshev, S.G.~Novokshonov
\footnote{e-mail: nov@otf.fti.udmurtia.su}\\
Physics--Technical Institute, Ural Division of RAS\\
Kirov str., 132, Izhevsk, 426001, Russia}
\date{}
\maketitle

\begin{abstract}
The spatial non--locallity (dispersion) of the transport equations results
in a nonli\-near dependence of the voltage drop $U$ on the distance between
the points of measuring. Therefore the results of the usual two--probe
measurements of the conductivity depend essentially on the relation between
the sample linear size $L$ and the spatial dispersion scale $R$ of the
generalized diffusion coefficient $D(q,\omega)$. This makes it possible
to get information on the character of the spatial non--locality of
$D(q,\omega)$ in the Anderson localization regime and, in particular, on
the value of the correlation multifractal dimension $D_{2}$ of the electron
wave functions near the mobility edge.  
\end{abstract}
\leftline{PACS 71.30.+h, 72.15.Rn}
\bigskip
\large
\hspace*{6mm}{\bf 1.} In the last few years considerable attention has
been focussed on the problem of spatial dispersion (non--locality) of the
kinetic coefficients of dis\-ordered systems in the Anderson localization
regime \cite{Chalker-90,
BHS-96,Suslov-98,Suslov-95,GN-97,NG-98}. The cause
of such interest was the realization of the fact that the character of the
$q\omega$--dependence of the generalized diffusion coefficient $D(q,\omega)$
near the mobility edge is intima\-te\-ly connected to critical behavior
of the electron wave functions and, in the end, is determined by the scenario
of the metal--insulator transition \cite{Chalker-90,Suslov-98}. Indeed, the
Berezinskij\,--\,Gorkov localization criterion \cite{BG-79} requires that
in the localized phase $D(q,0)$ vanish simultaneously for all values of the
wave number $q$, and, according to the one--parameter scaling hypothesis,
the relation 
\begin{equation}
\label{Scaling}
D(t;q,\omega)=b^{2-d}D(b^{1/\nu}t;bq,b^{d}\omega)\,,\qquad 2<d<4
\end{equation}
must be satisfied \cite{AL-86}. Here $t=({\cal E}-{\cal E}_{c})/{\cal E}_{c}$
is the distance to the mobility edge ${\cal E}_{c}$, $b$ is a scaling factor,
$d$ is the space dimension, $\nu$ is the correlation length critical index. 

There are two qualitatively different versions of the critical behavior of
$D(q,\omega)$ that obey these general requirements. 
\medskip

$\bullet$ According to Chalker's hypothesis \cite{Chalker-90}, 
the multifractal structure of the electron wave functions near the
Anderson transition $(t\to 0$ and/or $\omega\to 0)$ leads to
an anomalously {\it strong} spatial dispersion of the generali\-zed
diffusion coefficient, whose scale is $R={\rm
 min}(\xi,L_{\omega})
\to\infty$, where $\xi\propto |t|^{-\nu}$ is the correlation length,
$L_{\omega}\propto\sqrt{D(\omega)/\omega}\propto
\omega^{-1/d}$ 
is the electron diffusion length in a time of $\sim
 1/\omega$ 
\cite{Chalker-90,BHS-96}. Depending on the relation between
$\xi,~L_{\omega}$ and $q$ Chalker \cite{Chalker-90} distinguishes
four main types of the asymptotic behavior:
\begin{equation}
\label{Chalker}
D(q,\omega)=D_{0}\left(\frac{l}{R}\right)^{d-2}\left\{\begin{array}{cc}
1\,, & qR\ll 1\,,\\

(qR)^{d-2-\eta}\,,~& qR\gg 1\,,
\end{array}\right.\,\qquad R={\rm
 min}(\xi,L_{\omega})\,.
\end{equation}
Here $D_{0}$ is the Drudian diffusion coefficient, $l$ is the mean free
path, $\eta$ is a critical index related to the correlation multifractal
dimension of the wave functions $D_{2}$ $(\eta=
d-D_{2})$ \cite{BHS-96}.

$\bullet$ On the other hand, Suslov's symmetry approach to the localization
theory \cite{Suslov-95} predicts {\it supression} of the spatial dispersion
of the diffusion coefficient in the vicinity of the Anderson transition
down to atomic scales $\sim\lambda_{F}$. More recently, this conclusion
was confirmed within the framework of a generalized formulation
\cite{GN-97,NG-98} of the self--consistent Vollhardt --- W\"olfle 
theory
\cite{VW-80}. Accor\-ding to
 \cite{GN-97,NG-98}, in the Anderson localization
regime
\begin{equation}
\label{NG-98}
D(q,\omega)=\frac{D(t,\omega)}{1+(qR)^{2}}\,,\qquad qR\ll 1,
\end{equation}
where the non--locality scale $R\propto\sqrt{D_{0}
\tau}|D(t,\omega)/D_{0}|$
and decreases as $D(t,\omega)\!\propto\big[{\rm
 min}(\xi,L_{\omega})
\big]^{2-d}$ until saturation is reached at $R\sim\lambda_{F}$.
\medskip

In the survey by Suslov \cite{Suslov-98} it is pointed out that the
absence of anomalously strong spatial dispersion of the generalized
diffusion coefficient near the mobility edge does not contradict the
concept of multifractality of the electron wave functions, it only
indicate that the equality $\eta=d-2$ (or $D_2=2$) should be satisfied.
The well--known Wegner's result $\eta=2\epsilon$ ($\epsilon=d-2\ll 1$)
\cite{Wegner-80} is directly connected with the critical behavior of
the inverse participation ratio. At the same time, the relation between
this quantity and $D(q,\omega)$ used in \cite{Chalker-90,BHS-96,SO-99}
cannot be considered correct for several reasons.\footnote{The detailed
discussion of the basic arguments pro and con hypothesis $\eta=d-2$ may
be found in \cite{Suslov-98}.} Therefore, in our opinion, the above
contradiction is only apparent. The same may be said about the results
of numerical modelling $\eta=1.2\pm 0.15$, $\eta=1.3\pm 0.2$, $\eta=1.5
\pm 0.3$ \cite{BHS-96} and $\eta=1.3 \pm 0.2$ \cite{SO-99} obtained by
different methods for $d=3$.  
 
This dilemma, touching upon the fundamental notions of the An\-der\-son
localization, calls for both theoretical and experimental solution. 
In this letter we derive a material equation relating the current density
in the spatially nonuniform case to the experimentally measured 
difference of the electrochemi\-cal potentials, and propose a measuring scheme
that allows one to obtain information on the degree of non--locality
of the diffusion coefficient of charge carriers. 
\bigskip

{\bf 2.} The voltage drop $U$ measured in the spatially nonuniform case 
is equal to the difference in electrochemical potential $\Delta U=\Delta
\zeta/e=
\Delta\varphi+\Delta\mu/e$ between the corresponding points of
a conductor. Therefore in calculating the current density it is necessary
to take into account the response of the system to both the mechanical
perturbation (electrical potential $\varphi$) and the thermal one
(chemical potential $\mu$) caused by the nonuniform electron
distribution induced in the conductor.\footnote{The nonuniform spatial 
electron distribution results not only in perturbation of the electrical
field in the conductor, which is involved in $\varphi$, but also in the 
appearance of the diffusion term in the measured total current density. 
Just the latter term is the response of the system to the thermal
perturbation.}    

Applying the general equations of the linear response theory
\cite{Zubarev-79,Kalashnikov-74} to the problem under consideration, we 
obtain a material equation relating the current density to the 
electrochemical potential gradient. for $q\ll k_{F}$ its Fourier
repre\-sen\-ta\-tion has the form  
\begin{equation}
\label{Matter-Eq}
j(q,\omega)=-iqen_{F}D(q,\omega)\zeta(q,\omega)=
-iq\sigma(q,\omega)U(q,\omega)/e\,,
\end{equation}
where $n_{F}$ is the density of states at the Fermi level, $\sigma(q,\omega)$ 
is the measured electrical con\-duc\-ti\-vi\-ty connected with $D(q,\omega)$
by the Einstein relation. It should be emphasized that $\sigma(q,\omega)$
is not Kubo's kinetic coefficient 
\begin{equation}
\label{Kubo-Cond}
\widetilde\sigma(q,\omega)=e^{2}n_{F}\frac{D(q,\omega)}{1+i(q^{2}/\omega)
D(q,\omega)}
\,,
\end{equation}
which, unlike $\sigma(q,\omega)$ (see Eq.~(\ref{Matter-Eq})), relates the
total current density to the electrical field acting in the system. It is
in the uniform case ($q=0$) only that $\sigma(0,\omega)=\widetilde\sigma
(0,\omega)$, and Eq.
~(\ref{Kubo-Cond}) coincides with the Einstein relation.

Eqs. 
(\ref{Matter-Eq}) and (\ref{Kubo-Cond}) describe the non--local linear
response of a spatially unboun\-ded and homogeneous system. In the general
case, the non--locality of material equations is of a more complicated
character. In the typical experimental situa\-tion, the sample has the shape
of a plane--parallel slab of thickness $L$. In this case the diffusion
propagator of charge carriers $\widetilde{G}(x,x';\omega)$ is a solution
of the equation
\begin{equation}
\label{Diffuson-Eq}
-i\omega\widetilde{G}(x,x';\omega)-\frac{\partial}{\partial x}
\int\limits_{-L/2}^{L/2}\widetilde{D}(x,y;\omega)\frac{\partial}
{\partial y}\widetilde{G}(y,x';\omega){\rm d}y=\delta(x-x')
\end{equation}
with the open boundary conditions
\begin{equation}
\label{Boundary-Condition}
\widetilde{G}(x,x';\omega)\vert_{x,x'=\pm L/2}=0\,.  
\end{equation}
The integral kernel of this equation (non--local diffusion coefficient)
enters into the material equation that relates the electrical current to
the electrochemical potential gradient (voltage drop)
\begin{equation}
\label{Nonuniform-Matter-Eq}
I(x)=-e^{2}n_{F}S\int\limits_{-L/2}^{L/2}\widetilde{D}(x,x';\omega)
\frac{\partial}{\partial x'} U(x'){\rm d}x'\,,
\end{equation}
where $S\propto L^{d-1}$ is the area of the sample cross section. The
solution of the boundary problem (\ref{Diffuson-Eq}),
(\ref{Boundary-Condition}) in the absence of spatial dispersion
$(\widetilde{D}(x,x';\omega)=D(\omega)\delta(x-x')\,)$ is given by
well--known method of images
\cite{Vladimirov-71}
\begin{eqnarray}
\label{Nonuniform-Propagator}
\widetilde{G}(x,x';\omega)\!&=&\!\sum_{n=-\infty}^{+\infty}\!\Big[G\big(x
-x'+2nL;\omega\big)-G\big(x+x'+(2n+1)L;\omega\big)\Big]\,,\nonumber\\[12pt]
G(x;\omega)&=&\ds\frac{1}{L}\sum_{q}\frac{\exp(iqx)}{-i\omega +q^{2}
D(\omega)}\,.
\end{eqnarray}
Here $G(x,\omega)$ is the diffusion propagator of charge carriers in a
spatially unboun\-ded homo\-ge\-neous system. We assume that the solution
(\ref{Nonuniform-Propagator}) holds true in the presence of spatial
dispersion if its scale is small in comparison with the sample size
$(R\ll L)$. In this case the integral kernel of the material equation 
(\ref{Nonuniform-Matter-Eq}) is expressed, by analogy with
(\ref{Nonuniform-Propagator}), in terms of the generalized diffusion
coefficient $D(q,\omega)$ of the spatially unbounded and homogeneous system
\begin{equation}
\label{Nonuniform-Kernel}
\widetilde{D}(x,x';\omega)\!=\!\frac{2}{L}\sum_{n=0}^{+\infty}\Big[D
(\overline q_{n},\omega)\sin\overline q_{n}x\,\sin\overline q_{n}x'+D
(q_{n},\omega)\cos q_{n}x\,\cos q_{n}x'\Big]\,,
\end{equation}
where $\overline q_{n}=2\pi n/L$ and $q_{n}=\pi(2n+1)/L$
 are discrete
values of the wave numbers.

Taking into account that the voltage drop $U(x)$ is an odd function of $x$,
and the current strength $I(x)=I$ is constant along the studied sample,
we will represent them through the corresponding Fourier series. Then,
upon substitution of (\ref{Nonuniform-Kernel}) in the material equation
(\ref{Nonuniform-Matter-Eq}), it is not difficult to find the expression
for the voltage drop Fourier coefficient $U_{n}$. Finally the
$x$--de\-pen\-den\-ce of the voltage drop can be presented as the
following Fourier series:         
\begin{equation}
\label{Delta-U(x)}
U(x)=\frac{4I}{LSe^{2}n_{F}}\sum_{n=0}^{+\infty}
\frac{(-1)^{n+1}}{q_{n}^{2}D(q_{n},\omega)}\,\sin q_{n}x\,,
\qquad |x|<L/2\,.
\end{equation}

\bigskip
{\bf 3.} When the diffusion coefficient is independent of the wave number
$q_{n}$ the series (\ref{Delta-U(x)}) gives, for a finite--size conductor,
the usual conductance definition  $g(L,\omega)=L^{d-2}e^{2}n_{F}D(0,\omega)
=L^{d-2}\sigma(\omega)$ and describes the linear $x$--dependence of the
voltage drop $U(x)$ within the sample $(|x|<L/2)$. The spatial non--locality
of the generalized diffusion coefficient $D(q_{n},\omega)$ changes the
conductance definition as:
\begin{equation}
\label{General-Coductance}
\frac{1}{g(L,\omega)}=\frac{8}{LSe^{2}n_{F}}\sum_{n=0}^{+\infty}
\frac{1}{q_{n}^{2}D(q_{n},\omega)}
\end{equation}
and results in a nonlinear $x$--dependence of the voltage drop
(\ref{Delta-U(x)}). Thus, infor\-ma\-tion on the spatial dispersion of the
generalized diffusion coefficient of charge carriers $D(q,\omega)$ can be
obtained by measuring the nonli\-near part of $U(x)$ (\ref{Delta-U(x)}):
\begin{equation}
\label{Measured-Signal}
\Delta U(x)=U(x)-\frac{2x}{L}U\big(L/2\big)\,.
\end{equation}

\begin{figure}[tbh]
\centering
\epsfxsize=8.25cm\epsffile{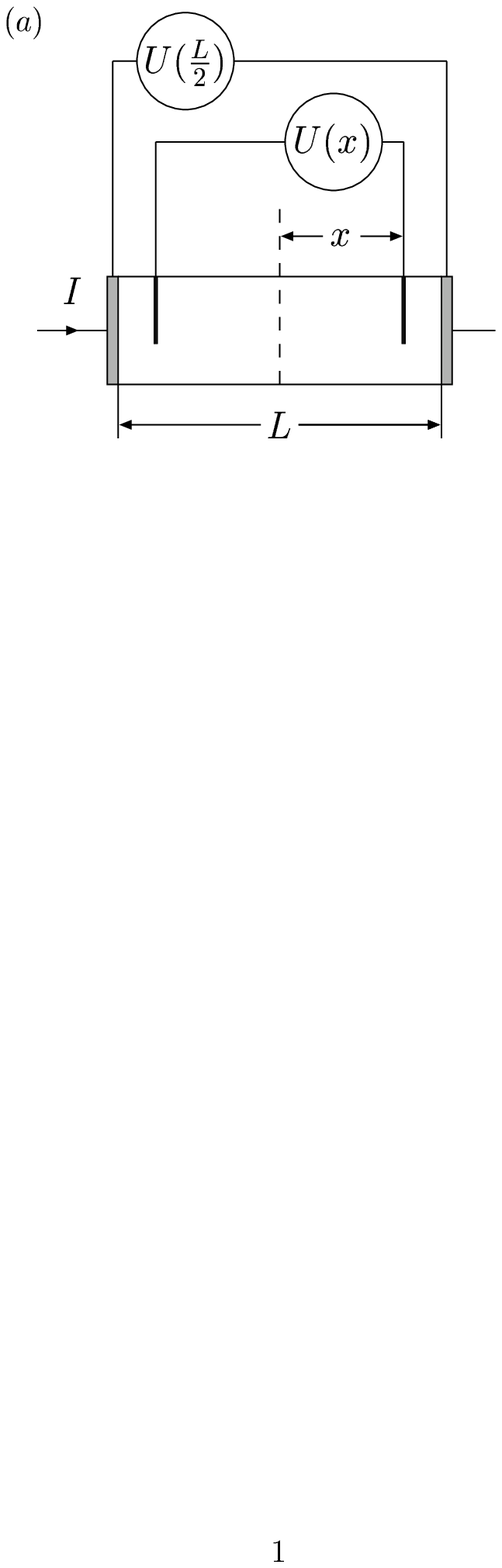}\epsfxsize=8.25cm\epsffile{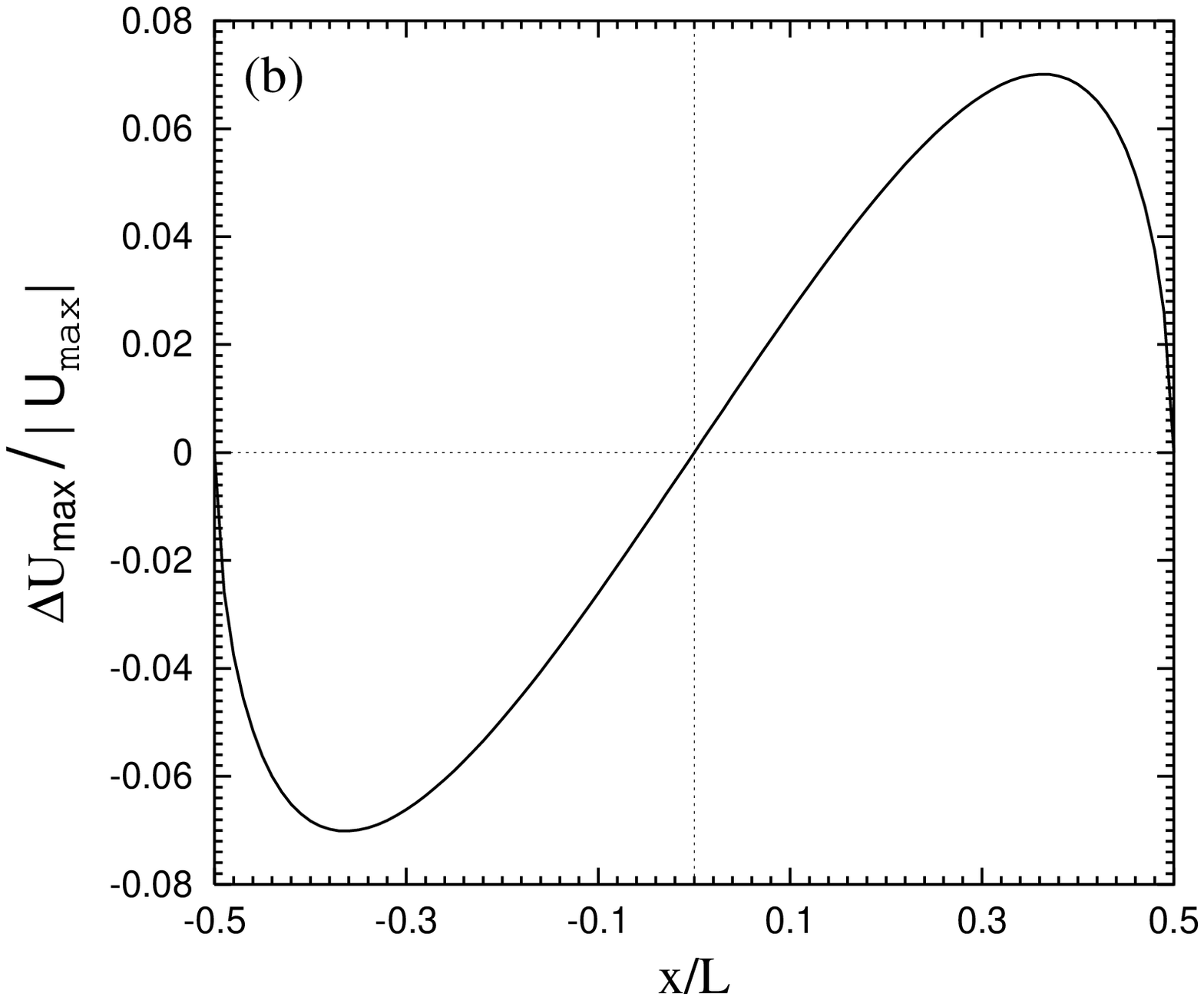}    
\caption{(a) The scheme of the measurement of the nonlinear part
of the voltage drop (here is depicted a longitudinal section of
investigated sample);
(b) Typical dependence of the nonlinear part of the voltage
drop on the potential probe position $x$. Curve is calculated using
Chalker's asymptotic formulas  (\ref{Chalker}) for $\eta=1.3$
and for $R/L=0.1$.}    
\label{fig:scheme}
\end{figure}

Consider a sample in the form of a plane--parallel slab of thickness
$L\gg R$ ($R$ is the spatial non--locality scale) with ideal ohmic contacts
on the opposite surfaces and with two potential measuring probes which
symmetrically lo\-cated at a distance $x$ from the mid--plane (see
Fig.~\ref{fig:scheme}(a)). It is best to situate the measuring probes near
the $x_{\rm max}$ points where $\Delta U(x)$ reaches its maximum value 
$\Delta U_{\rm max}$. Another pair of potential probes, the role of which
is played here by the ideal ohmic contacts, are necessary for measuring
the non--linear part of the voltage drop (\ref{Measured-Signal}).

Substituting (\ref{NG-98}) into (\ref{Delta-U(x)}), (\ref{Measured-Signal})
yields $\Delta U(x)=0$. Strictly speaking, this equality is fulfilled
in the case of ideal compensation of the linear part of the voltage drop. 
In any case, however, the measured signal is small in parameter $R/L\ll 1$. 
The nonzero contribution to $\Delta U(x)$ comes only from the nontrivial 
$q$--de\-pen\-den\-ce of the generalized diffusion coefficient from
(\ref{Chalker}). Fig.~\ref{fig:scheme}(b) shows typical dependences
of $\Delta U(x)$ calculated using linear interpolation between Chalker's
asymptotic formulas (\ref{Chalker}) for the inverse diffusion coefficient
$1/D(q,\omega)$ at $qR\ll 1$ and $qR\gg 1$. In this case $x_{\rm max}$ does
not depend on the scale of spatial non--locallity of $D(q,\omega)$ and
assumes values on the interval $x_{\rm max}\approx(0.70\div 0.76)L/2$ for
$\eta=1.1\div 1.5$. The non--linear part $\Delta U_{\rm max}$ calculated
at these points has the following asymptotic behaviors
\begin{equation}
\label{Measuring-Asymptotic}
\Delta U_{\rm max}\propto\left(\frac{R}{L}\right)^{\eta}\propto
\left\{\begin{array}{lc}
\omega^{-\eta/d}\,, & L_{\omega}\ll\xi\,, \\
\xi^{\eta}\propto|t|^{-\nu\eta}\,, & L_{\omega}\gg\xi\,,
\end{array}\right.\qquad L_{\omega}\,,\xi\ll L\,.
\end{equation}

Two ways for measuring the signal (\ref{Measuring-Asymptotic}) may be
suggested. The first is to analyze the frequency dependence of $\Delta
U_{\rm max}(\omega)$ in a sample with a fixed disorder level in a small
enough vicinity of the mobility edge on the metallic side of transition
$(|t|\ll 1)$. According to (\ref{Measuring-Asymptotic}) and the predictions
of Ref.~\cite{Chalker-90}, as the frequency decreases, an increase in
$\Delta U_{\rm max}\propto\omega^{-\eta/d}~~~(\omega\gg\omega_{c})$ should
be observed until saturation is reached at $\Delta U_{\rm max}\propto
(\xi/L)^{\eta}~~~(\omega\ll\omega_{c})$. The typical frequency $\omega_{c}$
is determined by the relation $L_{\omega_{c}}=\xi$ or $\hbar\omega_{c}
\approx \lambda_{F}^{-d}n_{F}^{-1}|t|^{\nu d}$, where $\lambda_{F}$ is
de Broglie's wave length at the Fermi level.

The second way is to measure the dependence of $\Delta U_{\rm max}$
(\ref{Measuring-Asymptotic}) on a dimension\-less distance $t$ to the
mobility edge ${\cal E}_{c}$ at a fixed frequency $\omega$. The 
stress--tuning tech\-ni\-que \cite{PSBR-86} seems to be the simplest
method of changing $t$ in the vicinity of ${\cal E}_{c}$. A very suitable
material for such measurements is Si:P, in which this technique makes
it possible to attain $|t|\approx 10^{-3}$ \cite{PSBR-86}.
According to (\ref{Measuring-Asymptotic}) and the predictions of Ref.~
\cite{Chalker-90}, as $t$ decreases, an increase in $\Delta U_{\rm max}
\propto t^{-\nu\eta}~~~(t\gg t_{\omega})$ should be observed until saturation
is reached at $\Delta U_{\rm max}\propto(L_{\omega}/L)^{\eta}~~~(t\ll
t_{\omega})$. Here the typical value $t_{\omega}$ is determined by the
relation $L_{\omega}=\xi$ or $\lambda_{F}^{d}n_{F}\hbar\omega\approx 
|t_{\omega}|^{\nu d}$.

\begin{figure}[tbh]
\centering
\epsfxsize=8.25cm\epsffile{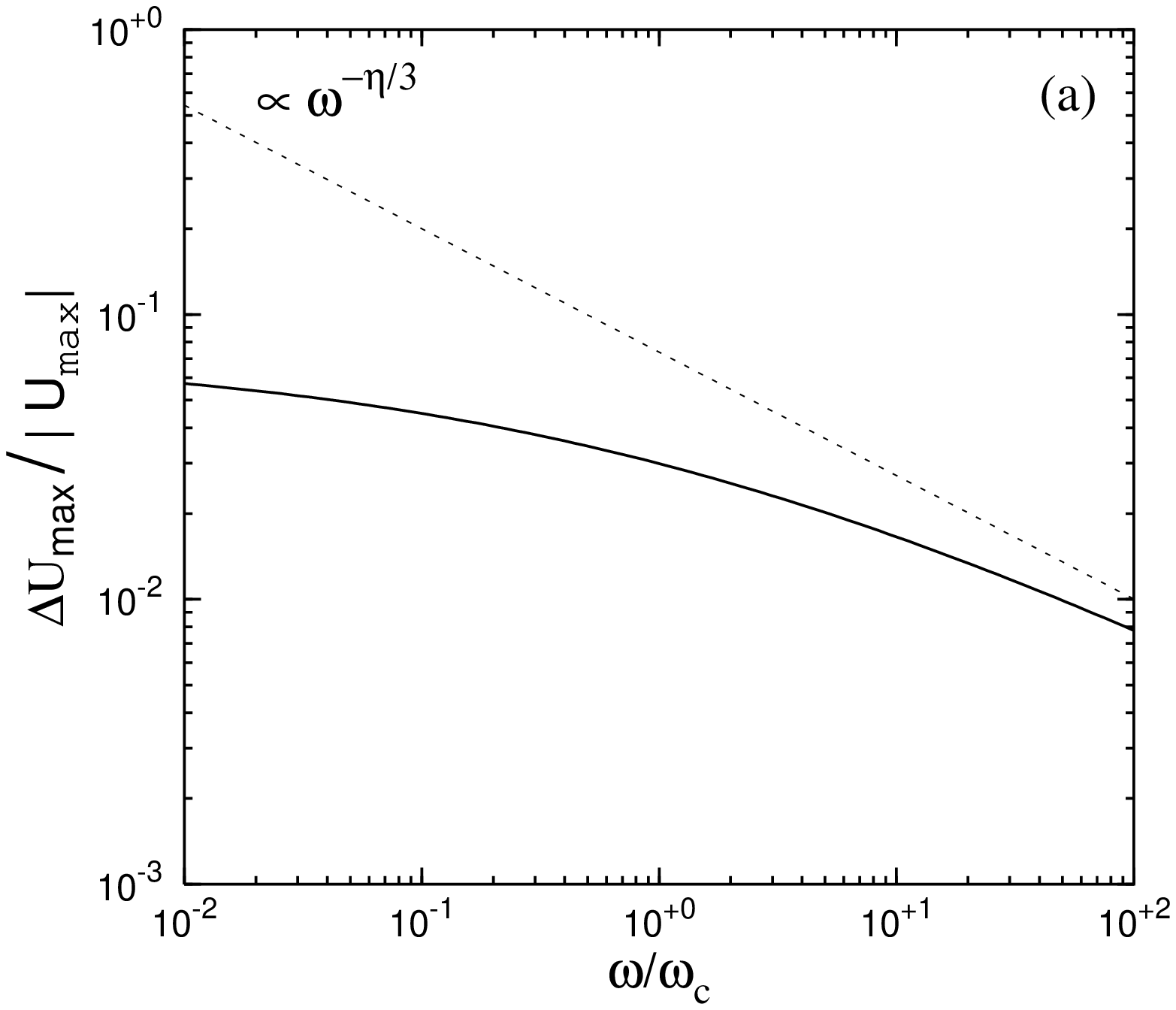}\epsfxsize=8.25cm\epsffile{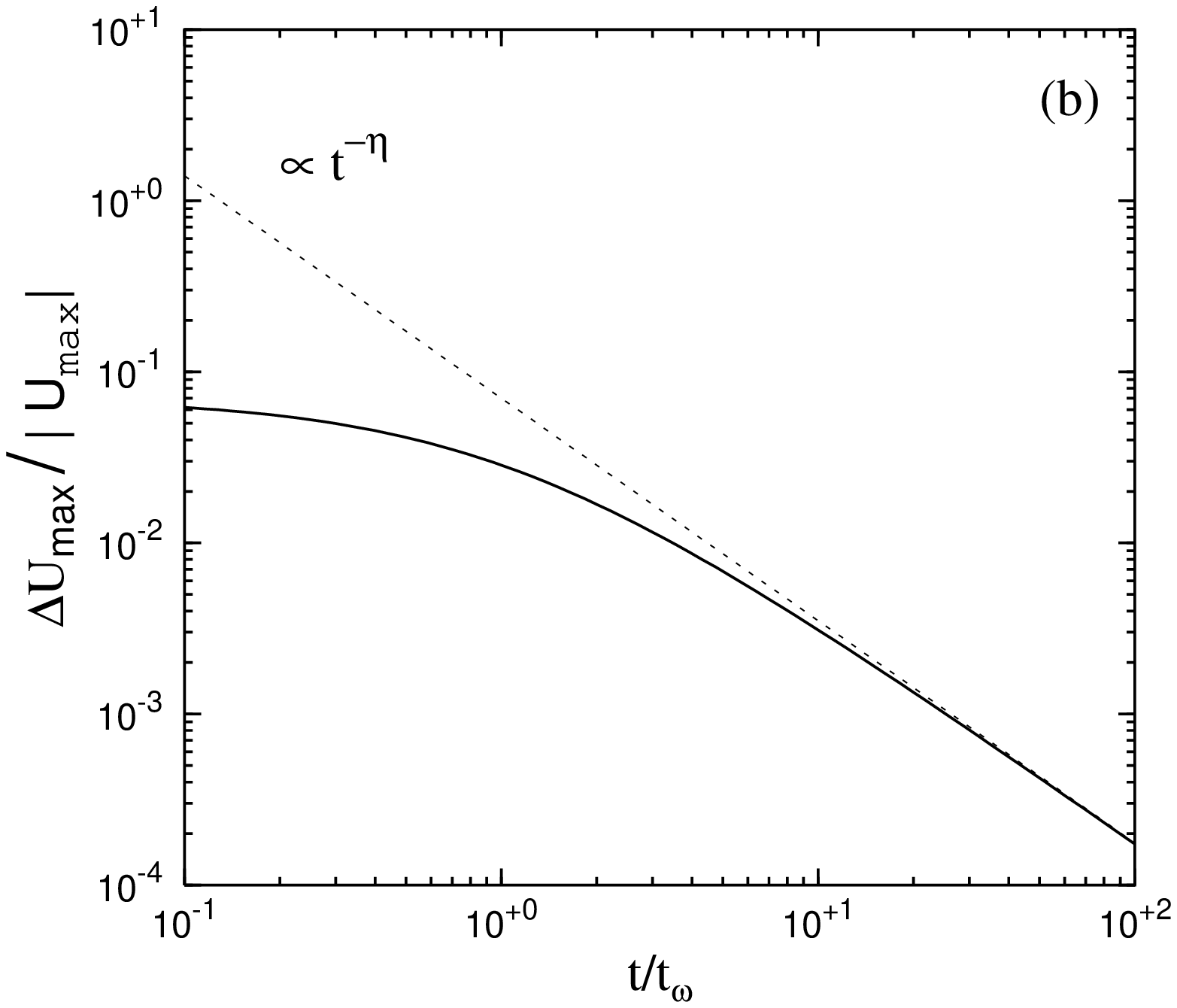}
\caption{Typical dependences of 
$\Delta U_{\rm max}$ on the frequency, $\omega$, (a) and dimensionless
distance from the mobility edge, $t$, (b) calculated by Eqs.
(\ref{Delta-U(x)}), (\ref{Measured-Signal}) using Chalker's expressions
(\ref{Chalker}) for $\eta=1.3$
. Dashed straight lines depict the asymptotic
behavior predicted by Eq. (\ref{Measuring-Asymptotic}).
}
\label{fig:plot}
\end{figure}

Such dependences obtained by Eqs.~(\ref{Delta-U(x)}), (\ref{Measured-Signal})  
using the interpolation $R^{-1}=\xi^{-1}+L_{\omega}^{-1}$ for the non--locality
scale, are plotted in Fig.~\ref{fig:plot}(a,b). Since in a sufficiently
small vicinity of the mobility edge the non--locality scale of $D(q,\omega)$
(\ref{Chalker}) takes on anomalously large values $R={\rm min}(L_{\omega},
\xi)$, the value of $\Delta U_{\rm max}$ (\ref{Measuring-Asymptotic})
is quite accessible for measurement. The estimates show that for Si:P
samples with the typical phosphorus concentrations $n_{\rm P}\approx
10^{18}\mbox{\rm cm}^{-3}$ and for really attainable values
$|t|\approx 10^{-2}\div 10^{-3}$, similar dependences of
$\Delta U_{\rm max}$ (see Eq.~(\ref{Measuring-Asymptotic})
and Fig.~\ref{fig:plot}) should be observed
in the frequency region accessible for probe measurements. For example,
at $t\approx 10^{-3}$ the typical frequence value is $\omega_{c}
\approx 10^{3}\div 10^{4}\mbox{\rm s}^{-1}$ (see Fig.~\ref{fig:plot}(a)), and the
corresponding correlation lengths are of order
 $L_{\omega_{c}}\sim\xi\sim 10^{-3}\div 10^{-2}\mbox{\rm cm}$.

So, we believe that the existence \cite{Chalker-90,BHS-96} or supression
\cite{Suslov-95,GN-97,NG-98} of the anomalous spatial dispersion of the
generalized diffusion coefficient near the Anderson transition can be
verified experimentally. 

\bigskip

The authors are grateful to I.M.~Suslov for having drawn their attention 
to this problem and for helpful discussions.

This work was supported by INTAS (Grant 99--1070).

\end{document}